\documentclass[conference]{IEEEtran}

\usepackage[dvipsnames]{xcolor}
\usepackage{tikz}
\usepackage{hyperref}
\usepackage{multirow}
\usepackage{cite}
\usepackage{amsmath}
\usepackage{comment}
\usepackage{colortbl}
\usepackage{multirow} 
\usepackage{array}

\ifCLASSINFOpdf
\else
\fi
\usepackage{url}

\usepackage[]{fancyhdr} % 
\usepackage{amsmath}
\usepackage{mathtools}
\newcommand{\changefont}{\fontsize{9}{9}\selectfont}
\IEEEoverridecommandlockouts
\begin{document}

%
% paper title
% Titles are generally capitalized except for words such as a, an, and, as,
% at, but, by, for, in, nor, of, on, or, the, to and up, which are usually
% not capitalized unless they are the first or last word of the title.
% Linebreaks \\ can be used within to get better formatting as desired.
% Do not put math or special symbols in the title.
\title{Distributed HVDC Emergency Power Control; case study Nordic Power System}

% author names and affiliations
% use a multiple column layout for up to three different
% affiliations
\author{\IEEEauthorblockN{Danilo Obradovi\'{c}, Mehrdad Ghandhari}
\IEEEauthorblockA{Division of Electric Power and Energy Systems\\
KTH Royal Institute of Technology\\
Stockholm, Sweden\\
\{daniloo, mehrdad\}@kth.se}
\and
\IEEEauthorblockN{Robert Eriksson}
\IEEEauthorblockA{Department of Power Systems\\
Svenska kraftn{\"a}t\\
Sundbyberg, Sweden\\
robert.eriksson@svk.se
}}

% conference papers do not typically use \thanks and this command
% is locked out in conference mode. If really needed, such as for
% the acknowledgment of grants, issue a \IEEEoverridecommandlockouts
% after \documentclass

% for over three affiliations, or if they all won't fit within the width
% of the page, use this alternative format:
% 
%\author{\IEEEauthorblockN{Michael Shell\IEEEauthorrefmark{1},
%Homer Simpson\IEEEauthorrefmark{2},
%James Kirk\IEEEauthorrefmark{3}, 
%Montgomery Scott\IEEEauthorrefmark{3} and
%Eldon Tyrell\IEEEauthorrefmark{4}}
%\IEEEauthorblockA{\IEEEauthorrefmark{1}School of Electrical and Computer Engineering\\
%Georgia Institute of Technology,
%Atlanta, Georgia 30332--0250\\ Email: see http://www.michaelshell.org/contact.html}
%\IEEEauthorblockA{\IEEEauthorrefmark{2}Twentieth Century Fox, Springfield, USA\\
%Email: homer@thesimpsons.com}
%\IEEEauthorblockA{\IEEEauthorrefmark{3}Starfleet Academy, San Francisco, California 96678-2391\\
%Telephone: (800) 555--1212, Fax: (888) 555--1212}
%\IEEEauthorblockA{\IEEEauthorrefmark{4}Tyrell Inc., 123 Replicant Street, Los Angeles, California 90210--4321}}

% <-this % stops a space

% use for special paper notices
%\IEEEspecialpapernotice{(Invited Paper)}

% The paper headers
%\lhead{11TH BULK POWER SYSTEMS DYNAMICS AND CONTROL SYMPOSIUM, JULY 25-30, 2022, BANFF, CANADA}
%\rhead{1}

%\fontfamily{phv}\fontseries{b}\fontsize{9}{11}\selectfont

% make the title area
\maketitle
\thispagestyle{fancy}
\pagestyle{fancy}

%\thispagestyle{fancy}
%\pagestyle{fancy}

% As a general rule, do not put math, special symbols or citations
% in the abstract
\begin{abstract}
Frequency Containment Reserves might be insufficient to provide an appropriate response in the presence of large disturbances and low inertia scenarios. As a solution, this work assesses the supplementary droop frequency-based Emergency Power Control (EPC) from HVDC interconnections, applied in the detailed Nordic Power System model.
EPC distribution and factors that determine the EPC performance of an HVDC link are the focus of interest. The main criteria are the maximum Instantaneous Frequency Deviation and used EPC power. The presented methodology is motivated based on the theoretical observation concerning linearized system representation. However, the assessed and proposed properties of interest, such as provided EPC active and reactive power, their ratio, and energy of total loads and losses in the system due to the EPC, concern highly nonlinear system behavior. Finally, based on the obtained study, remarks on the pragmatical importance of the EPC distribution to the frequency nadir limitation are provided.
\end{abstract}

\begin{IEEEkeywords}
\textcolor{black}{Control distribution, Frequency Containment Reserves, Emergency Power Control, HVDC power control, Nordic Power System}.
\end{IEEEkeywords}

% no keywords

% For peer review papers, you can put extra information on the cover
% page as needed:
% \ifCLASSOPTIONpeerreview
% \begin{center} \bfseries EDICS Category: 3-BBND \end{center}
% \fi
%
% For peerreview papers, this IEEEtran command inserts a page break and
% creates the second title. It will be ignored for other modes.
\IEEEpeerreviewmaketitle

\section{Introduction}

In the presence of today's trend of integrating renewable energy sources and increasing the capacity of High Voltage Direct Current (HVDC) interconnections, there is a natural consequence of a reduction in power system kinetic energy. In other words, system inertia tends to fall more often below values in which controlling frequency after a large disturbance becomes a significant challenge \cite{stab_clas}, \cite{FSI2}. In these cases, conventional Frequency Containment Reserves (FCR) provided by synchronous generators equipped by governor-turbine systems may be insufficient to provide adequate speed of response \cite{moj}, \cite{fcrd2}. Failing to do so, frequency may not be kept within the acceptable margins, and protection schemes are activated in the form of load shedding or generations tripping, leading to considerable costs.

Therefore, technical and economic motivations have already been recognized for introducing Fast Frequency Reserves (FFR) and power support over HVDC interconnections \cite{FFR}, \cite{ent_hvdc}. Depending on the system characteristics, the responsible System Operator (SO) defines compliant requirements towards reserves eg. FFR or HVDC systems. The essential properties of reserves are related to the capacity, control method or strategy, and market integration. Moreover, all of these aspects must be coordinated to bring an efficient and implementable solution. Procuring more reserves may decrease the frequency deviations. However, providing much more than needed implies an increased cost. Therefore, it is of interest to investigate the right amount of reserves and their speed of response to satisfy SO requirements.

%Therefore, it is of interest to investigate the right amount of reserves, together with their speed of response, needed to satisfy SOs requirements.

%Usually, in the assessment of these problems, distribution of reserves is seen mostly as a problem of reserve availability and their prices.

%%%%This work is focused on the problem of HVDC frequency support, where during a significant power imbalance in one synchronous area, through the change of active power reference on the connecting HVDC link, other area provides the required active power. 

%%%%However, it is also worthed to refer to other types of frequency support units since there is a methodology overlap in particular aspects.

%%%%Inertia reduction also jeopardizes the overall closed-loop stability margin of the system, which must be recognized and improved. Thus, the overall performance of the applied frequency service depends essentially on the implemented control method and the resources, which has to be carefully assessed.

%In the design of {frequency control}, literature tends to aggregate the same type of units, such as synchronous generators on one side and inverter-based resources (solar, wind, batteries, and HVDC links) on the other side. 
The total inertia and damping characteristic are considered unique system properties, vital for frequency control, and typically aimed to be augmented by inverter-based units. \cite{optDER}, \cite{Robert}. Thus, units' locations and the problem of distributing the reserves in the system are neglected. The primary reason is the assumption of a decoupling between active power-frequency and reactive power-voltage loops \cite{Kundur}. %This in order to be able to theoretically approach the design problem and to develop a generalised solution rather than a model dependent.   

More recent work \cite{VI1}-\cite{VIDmp3} opened the question of locations of the (fast) active power units in the context of their frequency control effectiveness. Studies \cite{VI1} and \cite{VI2} have shown how virtual inertia should be distributed among various locations in the grid to provide adequate support to the system in the presence of single or multiple disturbances. {Effectively, these studies analyzed the solutions to unequal distribution of mechanical inertia in large systems given in \cite{loc_inertia}. Further extensions of this problem are done in \cite{VIDmp1} - \cite{VIDmp3}, where the damping coefficient distribution is also considered.}

Comparing the different approaches, one can notice that the central aspect of the distribution issue is criteria that define satisfying response performance in terms of frequency deviation. An evaluation criterion is most often the maximum Instantaneous Frequency Deviation (IFD). In addition, studies discussing virtual inertia typically use the maximum Rate of Change of Frequency (RoCoF), which is important for some systems such as Great Britain or Ireland \cite{GBlow}, \cite{EirGrid}. Considering more factors in criteria as advanced studies in \cite{VIDmp1} and \cite{VIDmp3} do, such as the specific cost of power{/}energy delivered from the supporting units or frequency energy output to an impulse disturbance, bring more attention to the distribution matter. However, this does not necessarily imply the problem's urgency since, for SOs, control optimization may not be a priority, and many factors are hard to measure or penalize adequately.     

Even though recent studies opened the question of active power support distribution, they neglected the significant nonlinearities which follow the system response after a large disturbance. Those originate from the complex models of loads and other controllers, both frequency and voltage based, carefully designed and distributed in large systems impacting the overall system response.   
%%%Finally, the assessment where one explains more generally why one control unit performs better over the other in improving the frequency response into context of analyzed large system is missing. 

%The first one is reflected through the various activation levels among the utilized frequency control units.
%%%%However, the distribution matter might be overestimated in the cases where i) system is poorly small-signal stable (power system stabilizer (PSS) impact is totally neglected), ii) one considers only small disturbances and linearization, and iii) the goal is tight optimization rather than realistic system requirements which do not have to be so strict as optimization problem ie. pragmatical importance and real benefits are always quite important.

%those studies define the optimization method based on small disturbances and linearization. However, in those cases, there is a concern that the optimization method may not effectively distinguish the frequency and small-signal stability problem since, in real systems, Power System Stabilizers (PSS) usually provide sufficient damping. %Thus, it is still a question if frequency support distribution plays an important role in large disturbances in power systems where the maximum IFD is the main criteria.  

The analysis in{\cite{NEACDCPS}} illustrated the impact of voltage controllers and the importance of detailed large-scale systems for accurate FCR assessment. It is shown how excluding voltage dynamics, implying the change to total loads and losses response, can lead to incorrect IFDs values. Furthermore, besides directly providing additional active power, the studies \cite{svc1}-\cite{mbpss1} investigated the actual utilization of reactive power/voltage control for the frequency support purpose, while \cite{Milanovic} analyzed different load models' impact. %{Therefore, one needs to include the dynamic responses of loads and losses assessment into context of fast frequency control.} %%%%Therefore, it is important to include voltage dynamics and its nonlinearities, and rationale to assume that providing the same amount of active power from a different location will not reflect to the frequency response in the same way.

The focus of this work is the problem of distribution of Emergency Power Control (EPC) utilized in the HVDC interconnections. {The EPC presents a specific "emergency" control measure to support the FCR during severe disturbances (one form of FFR). Even though quite unlikely to occur, during these conditions, the system must respond adequately i.e., keep the maximum IFD within the required margins and avoid further disconnections of loads or generation.} %%%%%%So far, the impact of EPC distribution on the frequency response has not been studied in large systems. This paper contributes with a qualitative assessment of the placement of droop frequency-based EPC.

{EPC is applied as supplementary active power control loop in HVDC converter.} Depending on the type of converter, Line Commutated Converter (LCC) or Voltage Source Converter (VSC) type, one can notice different nature in dynamic active/reactive power response coupling \cite{Kundur}. {In other words, sometimes providing active power also implies the reactive power response.} These have not been studied before in the context of HVDC EPC, and their consideration is one of the contributions of this work. 

%Also, the operation conditions and control strategy impact the frequency response.

%The focus of this work is the problem of distribution of Emergency Power Control (EPC) utilized in the HVDC interconnections. So far, it is not specified or studied how exactly the EPC should be distributed and how specific distribution determines the EPC performance in large systems. This study investigates the droop frequency-based EPC referring to its confirmed performance provided in studies \cite{moj} and \cite{moj2}, and following proposals for the future EPC of the Nordic Power System (NPS) in \cite{ent_hvdc}.
%Moreover, depending on operating conditions, control strategies used in HVDC, and if a converter is Line Commutated Converter (LCC) or Voltage Source Converter (VSC) type, one can notice different nature in dynamic active/reactive power response coupling \cite{Kundur}. These have not been studied before in the HVDC frequency control context, and their consideration is one of the contributions of this work. 

%%%The unanswered question of previous work is \textit{how much the EPC distribution impacts the overall performance of that service?}

{This study aims to answer how the choice of HVDC link providing EPC, i.e. how its location in the system, impacts the maximum IFD. The approach investigates each HVDC link individually with the same EPC gain. The analyzed test case is large-scale Nordic Power System (NPS) experiencing highly nonlinear behavior after large disturbances. As the vital factors of interest, this paper is the first of its kind to address:}
\begin{itemize}
    \item {responses of both active and side-effected reactive power due to EPC from the chosen HVDC link;}
    \item {dynamic responses of total loads and losses implied by the EPC.}
\end{itemize}

{It is proposed how these factors should be considered and correlated with the EPC performance between various HVDC links. The theoretical motivation to support this methodology is also provided. Furthermore, the results of multiple disturbances is presented. Finally, the difference between efficient and inefficient distribution is illustrated through the needed total EPC active power.}

%%%The approach investigates each HVDC link individually. This is done by using the same EPC settings and assessing the improvement in maximum IFD. For that purpose, the EPC active power and as a consequence reactive power provided and responses of total loads and losses in the system are assessed. This paper is the first of its kind to provide how the analyzed factors should be considered and correlated with the EPC performance between various HVDC links. The theoretical motivation to support this methodology is also provided. Furthermore, the impact of multiple disturbances and their location is presented. The assessment is applied to the large-scale Nordic Power System (NPS) test case, given in \cite{NEACDCPS}, where the system is carefully modeled to capture the properties of interest.

%the same disturbance scenario and initial operating point, the aim is to study the factors affecting frequency improvement such as converter type, import/export scenario, EPC location to disturbance, or other system parameters vital for the given analysis. Then, the performance in improving the maximum IFD  Finally, specific distributions are tested to identify the exact EPC reserves needed to achieve the specific maximum IFD values. From here, the comparison between the "worst" and "best" are given, together with pragmatical remarks on EPC design and distribution in EPC. }

\section{Methodology}

The purpose of EPC is to support the frequency response after a significant power imbalance. That usually means the EPC activation comes with the specified frequency threshold, and reducing the maximum IFD is the main task of interest.~\footnote{Since there is the activation threshold, the maximum RoCoF is not affected.} Suppose there are many HVDC links potentially involved in the EPC service. In that case, the question is how EPC should be distributed and how significant that distribution is for the presented task? This work assesses this question in a case study with the NPS where, besides existing HVDC links, also the near-future links are involved. By studying this problem, one can understand how to use less power to provide more efficient frequency support for the investigated system. Less power also means less impact in the providing systems. %This study is focused on the NPS test case where most of the HVDC links are geographically in the South and South-West side of the system, and some of them are electrically close to each other, see Fig.~\ref{fig:sys_model}. 

Previous studies of EPC in the NPS in \cite{moj} and \cite{moj2} confirmed that droop frequency-based method, in overall, outperforms the currently applied ramp/step one. It is also shown that this type of HVDC supplementary power control improves the system small-signal stability \cite{DAN_PSS}, \cite{LH}. Furthermore, ENTSO-e and the respective SOs proposed in \cite{ent_hvdc} that frequency droop EPC should be considered for the future operation in the NPS. {That is why this study also utilizes the droop frequency-based EPC method. Nevertheless, it can be noted that one can apply the same methodology even for the other choice of the EPC design.} 

%\footnote{Even if there was no threshold, the impact would be minimal because the EPC method is droop frequency-based without virtual inertia part.}.

To analyze and compare the performance among different HVDC links' EPC, this study is interested in two aspects: i) the power responses of an HVDC link, and ii) respective system response investigated through total AC system loads and losses. The first one depends on the type of HVDC converter, its control modes, and operating condition. The latter is a function of the first, but also depends on the HVDC and disturbance locations. These relationships are explained in more detail in the following subsections. The analysis is given in the context of slower dynamics concerning system frequency control and converters' outer control loops.

\subsection{The relationships between HVDC active and reactive power responses}

\subsection*{LCC HVDC}

%According to \cite{Kundur}, reactive power consumption (LCC always consume the reactive power) can be defined as following:
%\begin{equation}
%    Q=P \tan{\phi},
%    \label{eq:Q_LCC}
%\end{equation}

According to \cite{Kundur}, reactive power consumption (LCC always consume reactive power) can be defined with the following:
\begin{equation}
\begin{aligned}
    Q&=P \tan{\phi},
    \label{eq:Q_LCC}
\end{aligned}
\end{equation}

where $P$ and $Q$ are active and reactive power of LCC, respectively, and the parameter $\phi$ is the angle by which the converter line current lags its voltage. Relationship \eqref{eq:Q_LCC} is valid for both rectifier and inverter, i.e., regardless if HVDC LCC is importing or exporting power, $P$ is positive since in the LCC power flow is predetermined.

In dynamic assessment, assuming that LCC uses the current set point to control the active power (as in the test cases of this work), then besides $P$, also $\tan{\phi}$ and $Q$ are affected \cite{Kundur}, as seen from:
\begin{equation}
    \begin{aligned}
    P&=V_dI_d;\\
    V_d&=V_{d0}\cos(\xi) -\frac{3}{\pi}X_cBI_d;\\ \cos(\phi)&=\frac{V_d}{V_{d0}},
    \label{eq:VdId_LCC}
\end{aligned}
\end{equation}
where $I_d$, $V_d$, and $V_{d0}$ are the "direct" properties of current, voltage, and no-load and zero firing angle voltage, respectively. The angle $\xi$ is either ignition delay ($\alpha$ for rectifier) or excitation advance ($\gamma$ for inverter) angle. Reactance $X_c$ reflect the commutation overlap phenomena and $B$ is the number of LCC bridges in series. More details are provided in \cite{Kundur}.

From here, it is observed that active and reactive LCC power control are coupled. The level of coupling depends on the operating point and LCC control design. 

%EPC injects supplementary positive active power into the system, assuming the under-frequency event. Referring to \eqref{eq:Q_LCC}, that implies two cases: 
%\begin{itemize}
%    \item LCC import - increase in absolute active power $P$, and increase in reactive power consummation $Q$.
%    \item LCC export - decrease in absolute active power $P$, and decrease in reactive power consumption $Q$.
%\end{itemize}

Even more, each LCC has reactive shunts switching on and off with predefined number of steps and power sizes. Their purpose is to support the voltage at Point of Common Coupling (PCC) and compensate for the reactive power consumption of an LCC. The activation is triggered at the specified LCC reactive power levels \cite{Kundur}. These ones are also taken into account when analyzing the EPC responses provided by LCC HVDC.~\footnote{In the system, there are also additional shunts to support the voltage stability regardless of the HVDC LCC links.}

%The dynamics of LCC shunts can be defined as:
%\begin{equation}
%    \dot{n}=\frac{1}{T_{sh}}(Q_{set}-Q_{meas}), \hspace{0.1 cm} 0 \leq n \leq n_{max}
%    \label{eq:shunt}
%\end{equation}

\subsection*{VSC HVDC}

%The reactive power of VSC HVDC can be decoupled from the active power and disturbances in grid when converter is set to reactive control mode. However, when VSC controls the AC voltage at the PCC, there is a reactive power response $Q_{vsc}$ due to voltage deviations in the system after a disturbance. 

%Additionally, that reactive power response further changes when there is an EPC utilized through the change of active power $Q_{vsc,\text{with EPC}}\neq Q_{vsc}$. Therefore 

The reactive power of VSC HVDC can be decoupled from its active power when the converter is set to reactive control mode. However, that is not the case when VSC controls the AC voltage at the PCC. That is because different active power injections affect the voltage magnitude at PCC. %%The extent of the impact is defined by system configuration and operating point. 

After a significant active power disturbance, besides frequency, voltage magnitudes are also disturbed. Voltage controllers installed in both converters and generators modify the reactive power output to reduce these oscillations. When EPC is additionally introduced, these reactive powers are further affected (since everything is coupled). 

These phenomena could be quite nonlinear and challenging to evaluate even more when the reactive current limits implemented in the converters are reached. Nevertheless, in further analysis, the effort is provided to consider those nonlinearities and recognize the properties of interest.

\subsection{The motivation from the linearized single-EPC single-frequency model}

First, the model is linearized in order to motivate which property adequately reflects the HVDC reactive power contribution due to the EPC. The system is assessed with the EPC from the $\text{i}^{th}$ HVDC link, and the EPC active power $\Delta p_{epc,i}$ is defined as:

%%To create a proper motivation and intuition which parameters or variables that are of interest the model is linearized.

\begin{equation}
    \Delta p_{epc,i}=-g'_i\Delta f,
    \label{eq:droopF}
\end{equation}
where $g'_i$ is the EPC gain (MW/Hz) and $\Delta f$ is the measured frequency deviation in the feedback loop. 

{Recognizing there is a certain coupling between active and reactive power in the HVDC control $\Delta q_{epc*,i}$\footnote{The star "*" represents the property implied due to only EPC{---}meaning that it is zero when there is no EPC.}, let this coupling be represented through $k_{pq,i}$ such that:}

\begin{equation}
\Delta q_{hvdc*,i}=k_{pq,i}\Delta p_{epc,i}.
\label{eq:dqepc}    
\end{equation}

{Both powers have positive direction of injection into the bus. Considering the control mode and HVDC type, it can be noticed that:
\[ 
  k_{pq,i} 
  \begin{dcases*} 
  =0 &  \text{VSC reactive control, or $Q_{hvdc}$ is limited} \\
  <0 &  \text{LCC import or VSC voltage control mode} \\ 
  >0 & \text{LCC export} 
  \end{dcases*} 
\]
}
%In the sense of linearized systems, $k_{pq,i}$ is a constant. However, that is not a case with nonlinear responses and that will be considered in the next subsection.

\begin{figure}[h]
\centering
\includegraphics[width=0.45\textwidth]{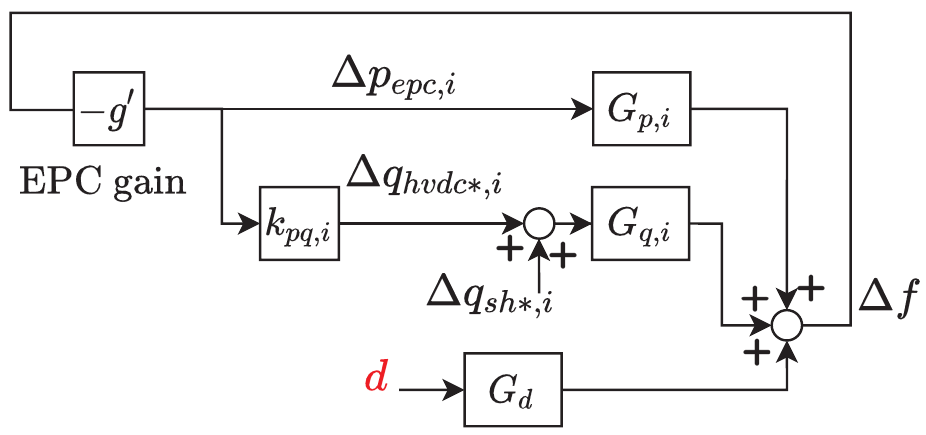}
\caption{Linearized system representation with the EPC illustrated for the $\text{i}^{th}$ HVDC link.}
\label{fig:Lin_sys}
\end{figure}

The closed loop system with a disturbance $d$ as input and frequency as output is shown in Fig.~\ref{fig:Lin_sys}.

The transfer functions $G_{p,i}$, $G_{q,i}$, and $G_{d}$ represent the input-output mapping between EPC active and reactive power, and disturbance (respectively) to frequency. The shunt reactive power $\Delta q_{sh*,i}$ is assessed as a separate input since it is not directly connected to the frequency. 

Then, the frequency can be represented through inputs $d$ and $\Delta q_{sh*,i}$ such as:

\begin{equation}
    \Delta f=\frac{1}{1+{g'G_{p,i}}+g'k_{pq,i}G_{q,i}}(dG_d+\Delta q_{sh*,i}G_{q,i}).
    \label{eq:df_d}
\end{equation}

%\begin{equation}
%    \Delta f=\frac{1}{1+\underbrace{g'G_{p,i}}_\text{EPC}+\underbrace{g'k_{pq,i}G_{q,i}}_\text{reactive side-effected}}(\underbrace{dG_d}_\text{disturbance}+\underbrace{\Delta q_{sh*,i}G_{q,i}}_\text{shunt input}).
%    \label{eq:df_d}
%\end{equation}

The special cases of this equation are:
\[ 
  \Delta f 
  \begin{dcases*}
  =\frac{dG_d}{1+g'G_{pi}+g'k_{pq,i}G_{q,i}} & \text{EPC, no shunt impact} \\
  =\frac{dG_d}{1+g'G_{pi}} &  \text{and no reac. coupling} \\
  =dG_d &  \text{no EPC = FCR-only} 
  \end{dcases*} 
\]

Further, \eqref{eq:df_d} can be reformulated in the following form:

%\begin{equation}
%\Delta f - \Delta p_{epc}G_{p.i}-\Delta q_{epc*,i}G_{q,i}-\Delta q_{sh*,i}G_{q,i}=dG_d,
%\end{equation}
%which is equivalent to:
\begin{equation}
    \underbrace{\Delta f}_\text{with EPC }  (1+g'(G_{p,i}+G_{q,i}\underbrace{(k_{pq,i}+\frac{\Delta q_{sh*,i}}{\Delta p_{epc}})}_{K_{pq,i}}))=\underbrace{dG_d}_{=\Delta f \text{ FCR-only}},
    \label{eq:df_ext}
\end{equation}

{where:
\begin{equation}
    K_{pq,i}=\frac{\Delta q_{hvdc*,i}+\Delta q_{sh*,i}}{\Delta p_{epc}}=\frac{\Delta q_{epc*,i}}{\Delta p_{epc}},
    \label{eq:df_ext2}
\end{equation}
}
{It is interesting to observe that $k_{pq,i}$, or equivalent, taking into account the shunt participation $K_{pq,i}$, impacts the frequency response. Therefore, the analysis should not focus on the side-effected reactive power alone but rather on its ratio with active power response due to EPC. In the following subsections, it is proposed how to also consider nonlinearities and their impact on the response.}% If HVDC links were connected at the same bus, this would be the only parameter determining the comparison among the frequency responses.}

{%However, the HVDC links are not connected to the same bus, and 
%%%The frequency deviations depend on the functions $G_{p,i}$ and $G_{q,i}$, which  depend on the system's overall design and location of the $i^{th}$ link.% The alternative is needed since these functions are hard to analyze (even in linear systems).
}

\textit{Remark:} {The previous observations are not based on rigorous analysis since the system is quite simplified, but they provide insights which variables could be of potential interest in the following studies related to the detailed nonlinear model representation.}

\subsection{Analysis of AC system loads and losses responses}

{As defined in \eqref{eq:droopF}, activation of active power EPC is in proportion to the frequency deviation i.e. less deviation means less EPC power provided for the same EPC gains. However, that information is not sufficient to explain what is happening with the overall system response. Therefore, the proposal focuses on assessing the EPC impact on the total loads and losses in the system.}

The location of the EPC may matter more if the loads are voltage-dependent and the flows change affect the losses in the system. The reasoning is that power injections impact the (nearby) load voltage magnitudes, affecting the load power and then finally, the frequency response. For example, in under-frequency cases, their increase negatively impacts the efforts to reduce the frequency deviations. Activating the EPC in various locations reflects on the total active power of loads and losses differently. Therefore, there is an interest to correlate them to EPC performance to improve the frequency nadir.

It can be noted that even though these are challenging to assess in large systems, the total changes in loads due to EPC are typically condensed in the nearby voltage-dependent loads. Also, the response of losses can be correlated to disturbance and EPC locations by considering the power flow scenario in the AC system.

%The reasoning is when loads and losses change due to EPC activation, it affects the power balance in the system. For under-frequency cases (such as losing a generator), their increase negatively impacts the efforts to reduce the frequency deviations. Thus, providing the EPC from the area of large voltage-sensitive loads or too long electrical distance disturbance could reduce the performance of the EPC service.

% Firstly, the impact from the coupling between the EPC active and reactive power is considered with respect to the frequency nadir improvement. Additionally, the EPC location is studied through the total change of loads and losses since they are the most important factors reflecting the localized nature of the EPC. 

\subsection{Testing the individual HVDC EPC through simulations and assessing the overall response}

The initial scenario is a dimensioning incident with only FCR. In that scenario, there is no HVDC active power contribution. To compare, the performance is evaluated for each link, one activated at a time, in the same scenario with the same EPC gain $g'$ (MW/Hz). 

The idea is to compare assessment criteria with the EPC response for active and reactive power HVDC (plus shunt) contributions, changes in active power loads and losses, and frequency improvement. The focus is on the frequency response until the maximum IFD, since that is the primary property of interest. 

\begin{figure*}[h]
\centering
\includegraphics[width=1\textwidth]{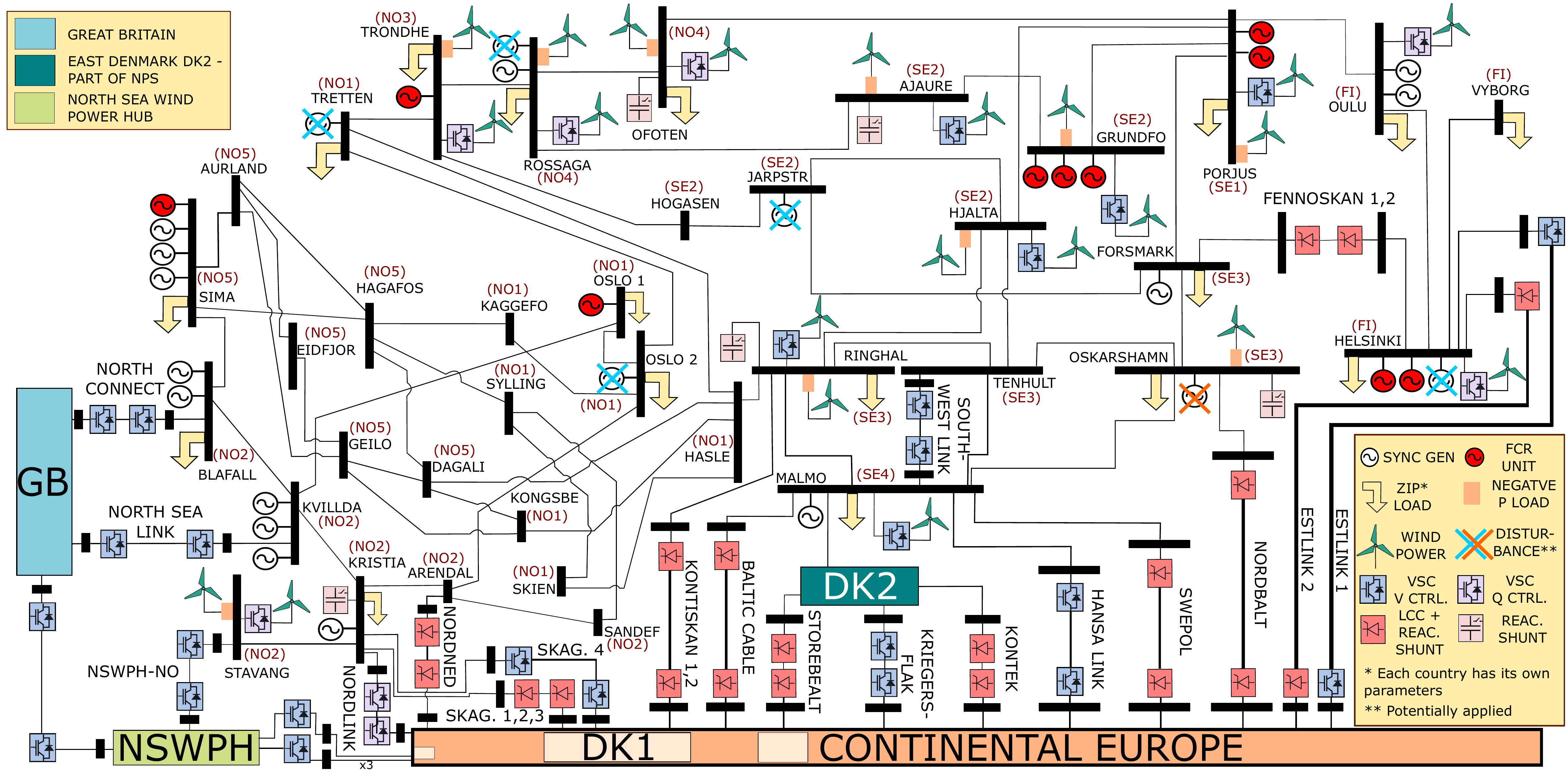}
\caption{Single line diagram/illustration of the Nordic Power System (NPS) test case with indicated elements of interest.}
\label{fig:sys_model}
\end{figure*}

%However, VSC converters controlling the AC voltage at the PCC (all HVDC and Swedish wind parks) provide reactive power support in order to control the voltage deviations affected by large disturbances. 

While testing the individual EPC performance of the $\text{i}^{th}$ HVDC link, the idea is to study the difference between cases (A) and (B) defined as:

\begin{itemize}
    \item \textit{(A)} FCR only, and
    \item \textit{(B)} FCR plus one EPC case of the $\text{i}^{th}$ HVDC link.\\
\end{itemize}
  
 The mechanisms behind the difference in performance are explained through reactive power impacting the loads and the transmission losses both impacted by the location of EPC support and disturbance.
The values and assessment criteria of interest are:
\begin{itemize}
    \item EPC active power injection:
        \begin{equation}
       \Delta P_{epc,i}= P_{\text{hvdc},B,i}
       -P_{\text{hvdc},A,i};
       \label{eq:P_delta}
    \end{equation}

    \item EPC reactive power change as a side effect of EPC:
    \begin{equation}
       \Delta Q_{epc*,i}= (Q_{\text{hvdc},B,i}+Q_{sh,B,i})
       - (Q_{\text{hvdc},A,i}+Q_{sh,A,i}),
       \label{eq:Q_delta}
    \end{equation}
    where $Q_{\text{hvdc/sh},A/B,i}$ is an inherent $\text{i}^{th}$ HVDC/ shunt (in case of LCC) reactive power response;

    \item The integral of a ratio between total reactive power change due to EPC and EPC active power support until the time of the frequency minimum $t_{min}$:
    \begin{equation}
    K_{pq\sum,i}=\int_{0}^{t_\text{min}}\frac{\Delta Q_{epc*,i}}{\Delta P_{\text{epc},i}} dt=\int_{0}^{t_\text{min}} K_{pq,i} dt;
        \label{eq:Kpqsum}
    \end{equation}
    The rationale is that the IFD is a function of the active power imbalance during the time from the disturbance to the IFD. Here, \eqref{eq:Kpqsum} captures the coupling of the reactive and active power hence the impact from reactive power.
    \item Total system load energy change due to EPC injection $\Delta E_{L,epc*}$ until the time of frequency minimum $t_\text{min}$, defined as:
    \begin{equation}
    \Delta E_{L,epc*}=\int_{0}^{t_\text{min}}\Delta P_{L,epc*} dt,
        \label{eq:EPL}
    \end{equation}
    where:
    \begin{equation}
       \Delta P_{L,epc*}= P_{L,tot,B}- P_{L,tot,A},
       \label{eq:PL_delta}
    \end{equation}
    and $P_{L,tot,A/B}$ is a total system active power load response;
 The rationale in \eqref{eq:EPL} is the change in loads in the time from the disturbance to the IFD.
    \item Total system losses energy change due to EPC injection $\Delta E_{\gamma,epc*}$ until the time of frequency minimum $t_\text{min}$, defined as:
    \begin{equation}
    \Delta E_{\gamma,epc*}=\int_{0}^{t_\text{min}}\Delta P_{\gamma,epc*} dt,
        \label{eq:EPgamma}
    \end{equation}
    where:
    \begin{equation}
       \Delta P_{\gamma,epc*}= P_{\gamma,tot,B}- P_{\gamma,tot,A},
       \label{eq:Pgamma_delta}
    \end{equation}
    and $P_{\gamma,tot,A/B}$ is a total system active power losses response;
    
    \item The improvement in the maximum IFD (nadir) due to EPC, obtained as:
    \begin{equation}
    \Delta f_{epc*}=\min(f_\text{B})-\min(f_\text{A}).
    \label{eq:imprv_ifd}
    \end{equation}
\end{itemize}

Assessing each link identifies the size of each factor and how they differ. Finally, the comparison among HVDC links performance is given for the various disturbance sizes and locations.   %Finally, various combinations of EPC distributions are tested, and frequency improvement is studied for different total EPC gain (MW/Hz) (or total EPC power in MW).\\

\textit{Remark:} It can be noted that there is no strictly fair and overall comparison between the HVDC links regarding the EPC since one would need to keep the same operating point and isolate the impact of the active and reactive power, which are coupled. That is not happening in the actual case with constant operating point change, various disturbances, and all nonlinearities. However, the goal here is to analyze the dominant factors in the EPC performance and specific characteristics for the given scenario and NPS test case.

\section{System model}

The studies performed in this work are applied to the Northern European AC/DC power system test case, {with the full system description regarding the NPS, HVDC links, neighboring system equivalents, and the North Sea Wind Power Hub (NSWPH) given in \cite{NEACDCPS}. Those descriptions relate to the modeling details, which are carefully chosen to serve the studies of frequency control assessment in the current and near-future scenarios for the NPS. This manuscript provides only the changes to the model given in \cite{NEACDCPS} and the most essential properties needed to follow the context of work. Since all components are represented by generic models and no proprietary information is involved, the model has been made publicly available and can be accessed at \cite{Git} and used through the DigSilent PowerFactory 2020 software \cite{PF2020} (or newer).}
%\begin{itemize}
%    \item NPS description concerning the typical loading and inertia values;
%    \item Frequency control division to different time scales and range of activation;
%\end{itemize}

{The single line diagram of the NPS part is given in Fig.~\ref{fig:sys_model}. The figure includes the legend with the essential elements used in the illustration. Since the study analyzes the NPS, the single line diagram comprises Finland, Sweden, Norway, and HVDC interconnections. DK2 (part of NPS) is simplified due to its smaller geographical and system size, and large complexity.}

The focus of interest is the NPS response for the largest N-1 power disturbance, the trip of nuclear power plant Oskarshamn 3, marked with an orange "X" sign over a generator in Fig~\ref{fig:sys_model}. This is a dimensioning incident for the FCR study in the NPS, and it creates a power step disturbance of 1450~MW and 103~MVAr with a kinetic energy loss of 3.29~GWs. The response constraints are defined such that the maximum allowed IFD and Steady-State Frequency Deviation (SSFD) are 0.9~Hz and 0.4~Hz, respectively \cite{NEACDCPS}. Other disturbances, indicated with the blue "X" signs in Fig.~\ref{fig:sys_model}, are also analyzed.

%%%%The focus of interest is the NPS response for the largest N-1 power disturbance, which is the trip of nuclear power plant Oskarshamn 3, marked with orange "X" sign over a generator in Fig.~\ref{fig:sys_model}. This is considered as a dimensioning incident for the FCR study in the NPS, and it creates a power step disturbance of 1450~MW and 103~MVAr, and kinetic energy loss of 3.29~GWs. The response constraints are defined such that the maximum allowed IFD is 0.9~Hz, and maximum allowed Steady-State Frequency Deviation (SSFD) is 0.4~Hz \cite{NEACDCPS}. Additionally, other large disturbances are analyzed and they are indicated with the blue "X" signs in Fig.~\ref{fig:sys_model}.

The FCR units, mainly responsible for taking care of a large disturbance, are distributed among ten synchronous generators marked red in Fig.~\ref{fig:sys_model}. The total regulating strength of the FCR is 3468~MW/Hz, designed to ensure the SSFD is below 0.4~Hz. Considering the typical turbine dynamics of FCR units in the NPS, the dynamic parameters of the governors are tuned to fulfill the exiting FCR dynamic requirements presented in \cite{fcrd2}.

%%%%The FCR units, mainly responsible for taking care of a large disturbance, are distributed among ten synchronous generators marked with red color in Fig.~\ref{fig:sys_model}. The total regulating strength of the FCR units is 3468 MW/Hz, designed to ensure the SSFD below 0.4~Hz. Considering the typical turbine dynamics of FCR units in the NPS, the dynamic parameters of the governors are tuned to fulfill the exiting FCR dynamic requirements, presented in \cite{fcrd2}. %Even though FCR turbines initially have various loading, the effective water time constant is equal among them and it is 1.2~s, referring to the range proposed in [].

To highlight the challenges in the frequency control, the inertia constant of non-hydro generation is reduced by half compared to the model given in \cite{NEACDCPS}~\footnote{There is no clear guideline on which nuclear or fossil-fuel units will be decommissioned in the future (apart from the ones already removed). Therefore, the alternative way of assessing lower inertia values was to reduce generator inertia constants of all non-hydro units.}. In that way, the total kinetic energy of the system is $E_k=$104.8~GWs, which is slightly smaller than the minimum inertia scenario ever recorded in the NPS \cite{inertia_scenario} - 110~GWs. Considering that the rated power of synchronous machines is $S_b=$36.9~GW, the system inertia constant is then $H=E_k/S_b=2.84$~s. It can be observed that now, FCR alone cannot keep the maximum IFD within the allowed margin of 0.9~Hz, and additional support from the EPC is required.

The test system has both current and future LCC and VSC HVDC interconnections that can be utilized for the EPC (all of them besides the Kriegers Flak). The operating point was adjusted according to the commonly observed power flow orientations in HVDC links and load centers to consider a plausible scenario for the future system. Consequently, HVDC links do not have the same headroom for providing the EPC, as shown in Table~\ref{tab:HVDCall}. $P_{r,i}$ represents the rated power of the link\footnote{In reality, the future NSWPH-NO link's capacity is still uncertain, and it might be set/decreased to 1400~MW.}, while $P_{hvdc,i}$ is its import active power.

%Authors have been informed that, after the new studies of respective TSOs, the future NSWPH-NO link's capacity might be set to 1400~MW.

%%%The test system has both current and future LCC and VSC HVDC interconnections that can be utilized for the EPC (all of them besides the Kriegers Flak). The operating point was adjusted according to the commonly observed power flow orientations in HVDC interconnections and load centers, so that a plausible scenario for the future system is considered. Considering this scenario, and various capacity ratings, HVDC links do not have the same headroom for providing the EPC, as can be observed from Table~\ref{tab:HVDCall}. $P_{r,i}$ represents the rated power of the link, while $P_{hvdc,i}$ is its import active power reference.

\begin{table}[ht!]
    \centering
    \caption{HVDC links potentially utilized for the EPC (the set of $E$).}
    \begin{tabular}{c c c c c c}
    \hline
    \hline
    \multirow{2}{*}{\#} & \multirow{2}{*}{Name} & \multirow{2}{*}{Acronym} & \multirow{2}{*}{Type} & {$P_{r,i}$} & {$P_{hvdc,i}$ }  \\ 
     &  &  &  & (MW) & (MW)   \\ 
        \hline
        1 & Baltic Cable & BC & LCC & 600 & 276  \\ 
        %\hline
        2 & Estlink 2 & EST2 & LCC & 650 & -468  \\ 
        %\hline
        3 & Kontek & K & LCC & 600 & 350  \\ 
        %\hline
        4 & Kontiskan 1 & KS1 & LCC & 380 & 200  \\ 
        %\hline
        5 & Kontiskan 2 & KS2 & LCC & 360 & 180  \\ 
        %\hline
        6 & NorNed & NoNd & LCC & 700 & -450  \\ 
        %\hline
        7 & Storebaelt & SB & LCC & 600 & 100  \\ 
        %\hline
        8 & Skagerrak 3 & SK3 & LCC & 500 & -350  \\ 
        %\hline
        9 & SwePol & SwPl & LCC & 600 & 100  \\ 
        %\hline
        10 & Skagerrak 1-2 & SK12 & LCC & 500 & -250  \\ 
        %\hline
        11 & NordBalt & NB & VSC & 700 & -600  \\ 
        %\hline
        12 & Estlink 1 & EST1 & VSC & 350 & -251  \\ 
        %\hline
        13 & Skagerrak 4 & SK4 & VSC & 700 & -250  \\ 
        %\hline
        14 & Hansa link & HL & VSC & 700 & 400  \\ 
        %\hline
        15 & Nord Link & NL & VSC & 1400 & -400  \\ 
        %\hline
        16 & North Connect & NC & VSC & 1400 & -487  \\ 
        %\hline
        17 & North Sea Link & NSL & VSC & 1400 & -787  \\
        %\hline
     18 & NSWPH-NO &	NSWPH &	VSC & 2100 &-443 \\
    
    \hline
    \hline
    \label{tab:HVDCall}
    \end{tabular}
\end{table}
  
%\footnote{Authors have been informed that, after the new TSOs testing, the future link's capacity might be set to 1400~MW.  
  
%The applied EPC method is droop frequency-based with a frequency threshold of 0.4 Hz. Each HVDC link measures its local frequency and, when activated, provides additional support by changing the active power reference by such as the following.

The applied EPC method is droop frequency-based with a frequency threshold of $f_\text{activ.,epc}=-0.4$~Hz. Each HVDC link measures its local frequency $\Delta f$~(Hz) and, when activated, provides additional support by changing the active power reference by $\Delta P_{epc,i}$~(MW) such as:

\begin{equation}
    \Delta P_{epc,i}=-g_i\frac{P_{r,i}}{f_n}\Delta f=-g'_i\Delta f,
    \label{eq:droopF2}
\end{equation}

where $g_i/g'_i$~(pu)/(MW/Hz) is the EPC gain and $f_n=50$~Hz is nominal frequency.

As shown in \cite{NEACDCPS}, depending on specific voltage control distribution in the system, there is a different maximum IFD. That is why it is necessary to carefully design the reactive power/voltage control applied in both generators and converters. All the generators illustrated in Fig.~\ref{fig:sys_model} and those with rated power above 40~MVA have implemented automatic voltage regulators and Power System Stabilizers (PSSs) to improve rotor angle stability.

%%%The generation units with nominal power above 40 MVA are represented behind their step-up transformers. All synchronous machines have a $5^{th}$ or $6^{th}$ order model (equivalent to model 2.1 and 2.2 according to \cite{IEEEgen}), including saturation effects, supplemented with automatic voltage regulator and Power System Stabilizer (PSS) models. The latter were tuned to ensure a sufficient level of rotor angle stability. The generators below 40~MVA were represented as negative loads.

An estimated onshore and offshore wind integration is included through VSC converters and negative active power loads. Following the current standards in the NPS, the "wind" VSCs integrated in Norway, Finland and Denmark control the reactive power. However, the ones installed in Sweden control the AC voltage to compensate for the loss of voltage control implied by the conventional generator decommission. Also, all VSC HVDC links control the AC voltage, besides Nord Link (NL) and NSWPH-NO. It is important to highlight that the reactive current of the VSCs has to be limited to 0.3~pu, and dynamic parameters controlling it are adjusted to avoid the bang-bang effect while providing voltage support during large disturbances.  

All LCC HVDC links have transformers with load tap changers and reactive shunts at both sides. Shunts have capacities equal to half of the converter's rated power. They are activated with five identical steps and a time constant of 0.5~s to support the voltage profile after faults or disturbances.

Loads are represented as static ZIP models with parameters obtained from the Nordic TSOs' study \cite{loads}. From here, it is noted that active power loads are voltage-dependent, having the possibility to impact the frequency response, which is vital for the analysis in this work.

%\footnote{Failing to recognize that the new adequate voltage control/reactive power support is needed during disturbances, leads to voltage instability.}

%\subsection{Remarks on small-signal stability and AC power flows impact}

%For the generic system models, it has been shown previously in \cite{DAN_PSS}, \cite{LH} that this type of control can also improve the small-signal stability. Studies in \cite{VIDmp1} and \cite{VIDmp2}, assuming second-order generator models, have shown that careful distribution of similar control supports (optimally) both frequency and small-signal system stability. However, in the existing and near-future NPS scenario, PSSs are capable of keeping the system damping to desirable values. Therefore, as long as EPC is not bringing apparent adverse side effects to system damping, small-signal stability is not included in the philosophy of EPC. However, it is to be acknowledged that locations of EPC support will have an impact on small-signal stability. 

%Another interesting matter is the EPC impact on the dynamic AC power flows. This question is not covered by the exact criteria, other than to avoid stressing further long corridors and getting at risk of transient instability.

\section{Results}

\begin{figure}[h]
\centering
\includegraphics[width=0.48\textwidth]{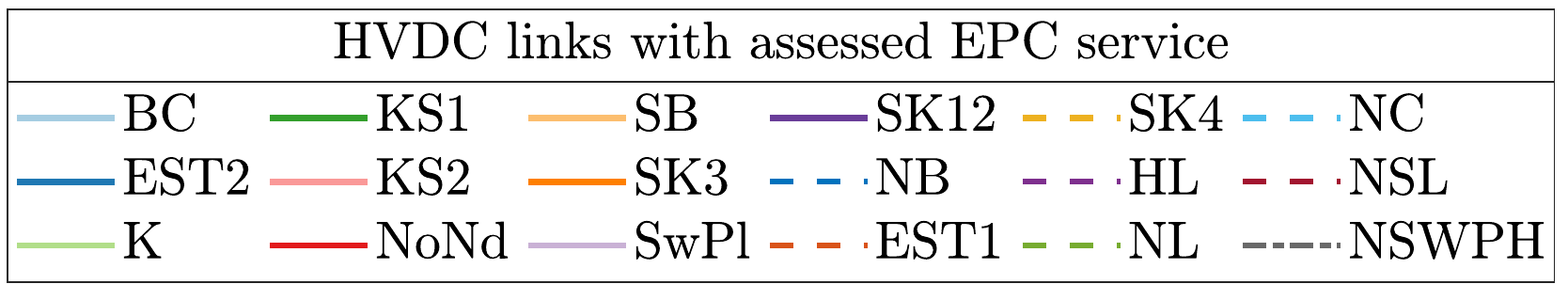}
\caption{List of HVDC links' acronyms which are individually tested for EPC support performance $\in E$. Used in the notation of future figures: LCC - solid lines, and VSC - dashed lines. Applied to figures 5, 7, and 9.}
\label{fig:HVDClist}
\end{figure}

In this section, the case studies illustrate the individual performance of each HVDC EPC support for the same MW/Hz gain in the presented NPS test case. The performance is assessed with respect to the improvement in maximum IFD, defined in \eqref{eq:imprv_ifd}. The acronyms of HVDC links tested for the EPC support are given in Fig.~\ref{fig:HVDClist} representing the set $E$, with the colors used for further plots and noticeable difference between marking LCC (solid lines) and VSC links (dashed lines). This analysis aims to answer how the distribution of EPC matters and the mechanisms behind. The performance is evaluated by the difference in the response between only FCR (case A) and EPC activated on one HVDC link (case B). 

\subsection{FCR response for the dimensioning incident}

\begin{figure}[h]
\centering
\includegraphics[width=0.49\textwidth]{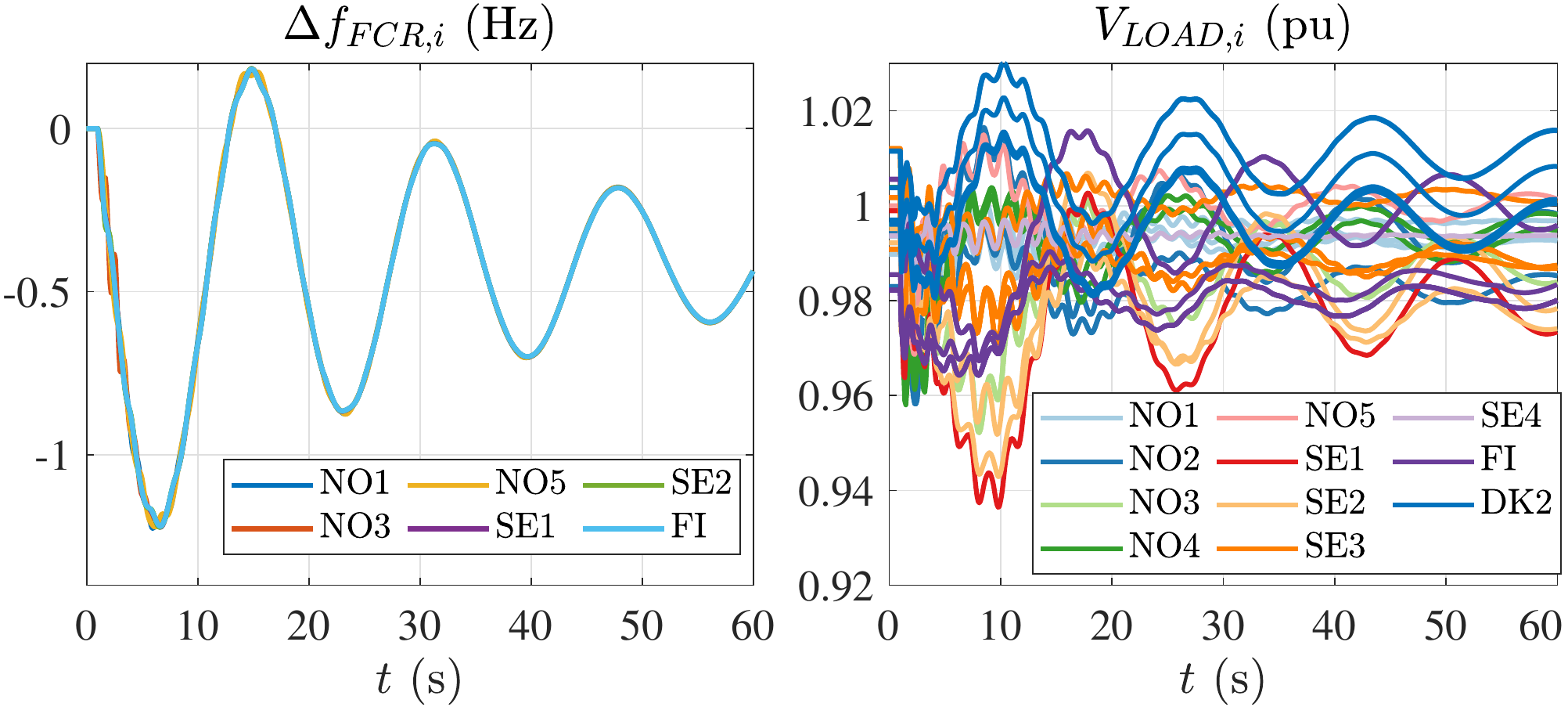}
\caption{FCR-only case (A) with responses of FCR frequency (on the left) and loads' voltage magnitudes (on the right) across the system. Different colors indicate the area where FCR unit or load is located (see Fig.~\ref{fig:sys_model}).}
\label{fig:FCR-only}
\end{figure}

Initially, the system response is presented in Fig.~\ref{fig:FCR-only} for the FCR-only case (A), where the applied disturbance is the trip of nuclear power plant Oskarshamn 3. Left plots in Fig.~\ref{fig:FCR-only} show the frequency response of ten FCR units distributed in the system and loads' voltage magnitude response in the whole system, on the right\footnote{DK2 area contains much more loads, but only four of them are given since there is a similar response from the rest of them.}. From these figures, it can be concluded that the system has a solid level of damping in-between frequencies but poor common-mode damping. That means PSSs provide sufficient support to system small-signal stability, while the low inertia implies large (system) frequency deviation. These oscillations can also be observed in voltage magnitude plots. Due to the more localized nature, voltage responses have various deviations across the system. They are a function of distributed system parameters and voltage control, which have to be carefully designed, especially for the cases of large disturbances.

\begin{figure}[h]
\centering
\includegraphics[width=0.48\textwidth]{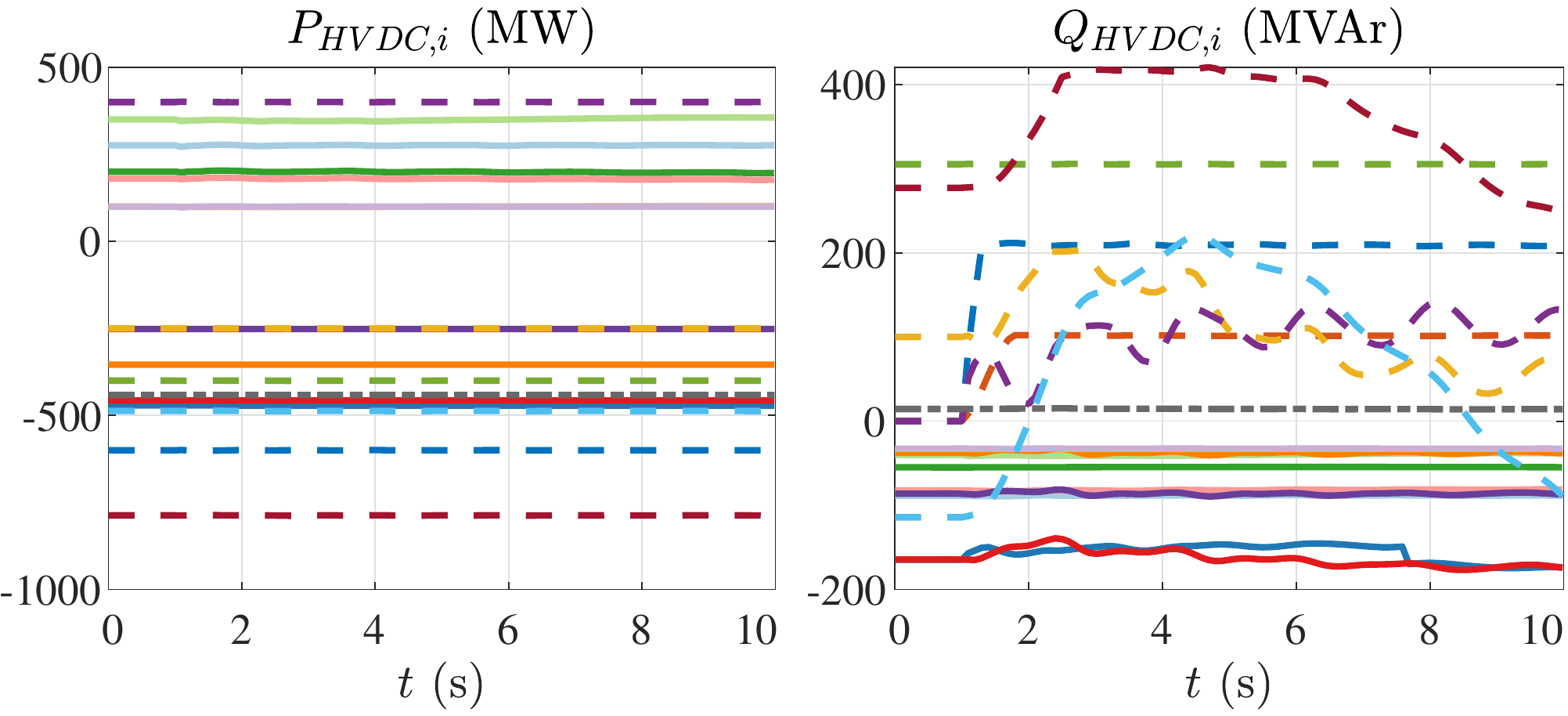}
\caption{FCR-only case (A) with responses of HVDC active and reactive powers with included shunts in case of LCC. Positive values indicate import or reactive power injection. See Fig.~\ref{fig:HVDClist} for legend.}
\label{fig:FCR_only_power}
\end{figure}

In the FCR-only scenario, the inherent responses of HVDC links' active (on the left) and reactive powers (on the right) are presented in Fig.~\ref{fig:FCR_only_power}. From here, one can see that active powers do not change (almost at all) since there is no supplementary action (i.e. no EPC). That is also the case for the reactive powers unless the VSC HVDC links control the AC voltage at PCC without reaching the current limit. That refers to a typical converter 0.3~pu reactive or 1~pu absolute current limit.

The responses of voltage and frequency must be stable and within the required limits, which is the case for voltage ($\pm 0.1$)~pu, but not for frequency where the maximum allowed frequency deviation is $\pm 0.9$~Hz. Therefore, the system needs EPC support, and further, it is studied how various HVDC links could reduce the maximum IFD.

\subsection{Individual EPC HVDC testing based on frequency improvement}

\begin{figure*}[h]
\centering
\includegraphics[width=0.8\textwidth]{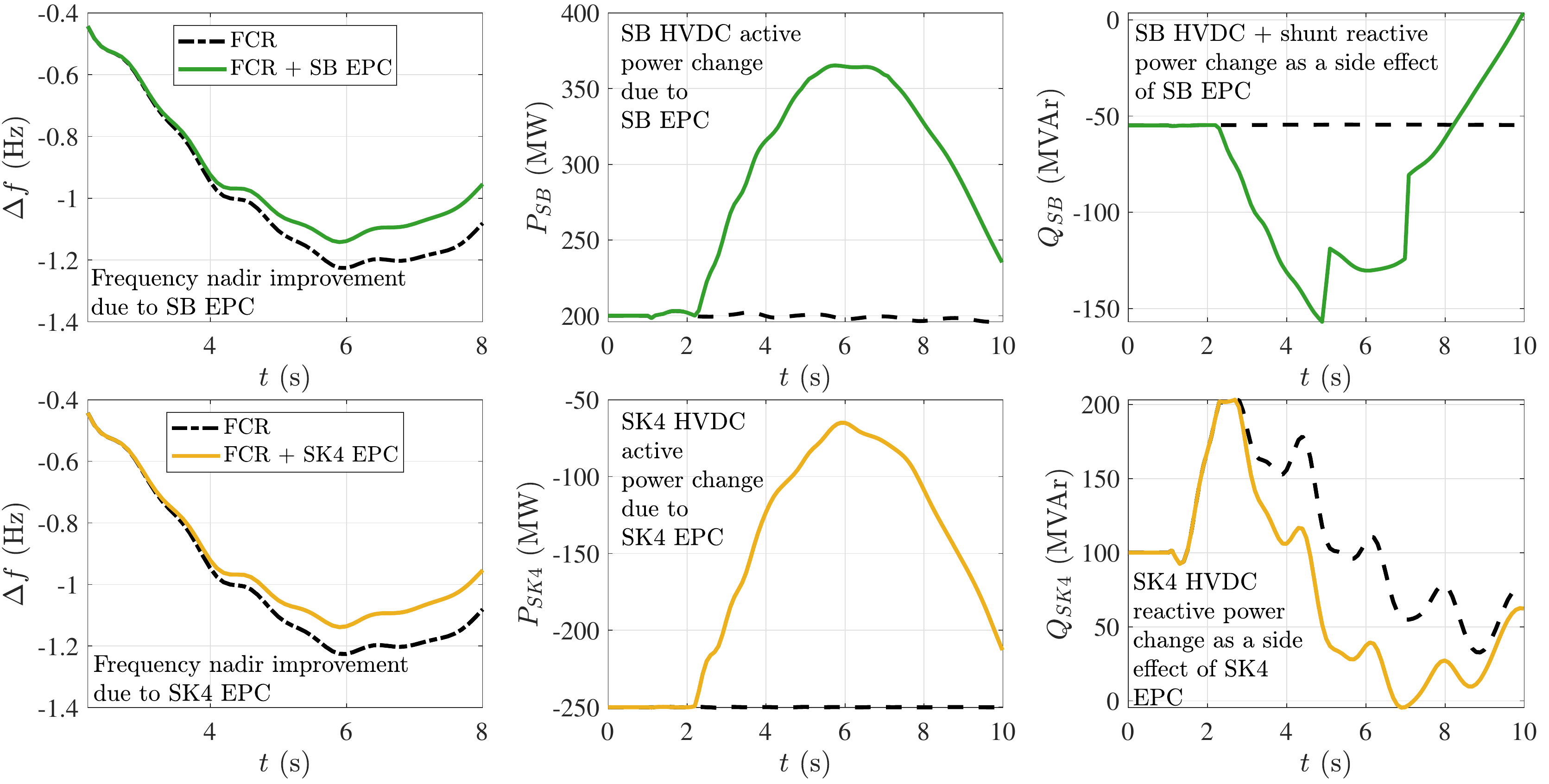}
\caption{System responses for testing two separate EPC support of StoreBealt (SB) and Skagerrak 4 (SK4) HVDC links for FCR-only (A) and EPC cases (B): uppers are SB EPC and bottoms are SK4 EPC. From left to right plots there are, respectively: frequency responses, and active and reactive power changes.}
\label{fig:SB_SK4_epc}
\end{figure*}

Each link is individually tested with the EPC gain equal to 250~MW/Hz, with the control defined by \eqref{eq:droopF2}. Two cases of HVDC links, LCC type StoreBealt (upper graphs related to SB) and VSC type Skagerrak 4 (bottom graphs related to SK4), are given in Fig~\ref{fig:SB_SK4_epc} to illustrate the contribution of individual EPC support.

Figure~\ref{fig:SB_SK4_epc} shows the system response comparison due to single EPC support for the frequency and active and reactive power change of the same link; graphs are presented from left to right, respectively. Active power change is a direct consequence of control defined in \eqref{eq:droopF2}, while reactive power change is a side effect resulting from control coupling. The main criterion of the single EPC performance is the ability to improve the maximum IFD (defined in \eqref{eq:imprv_ifd}), which is shown in the left graphs. The correlation to different improvement levels is partly explained through the change in the ratio between reactive and active power response behavior until the moment of nadir. %However, it is shown that the locations of both EPC and disturbance matter and have to be considered.

\begin{figure}[h]
\centering
\includegraphics[width=0.48\textwidth]{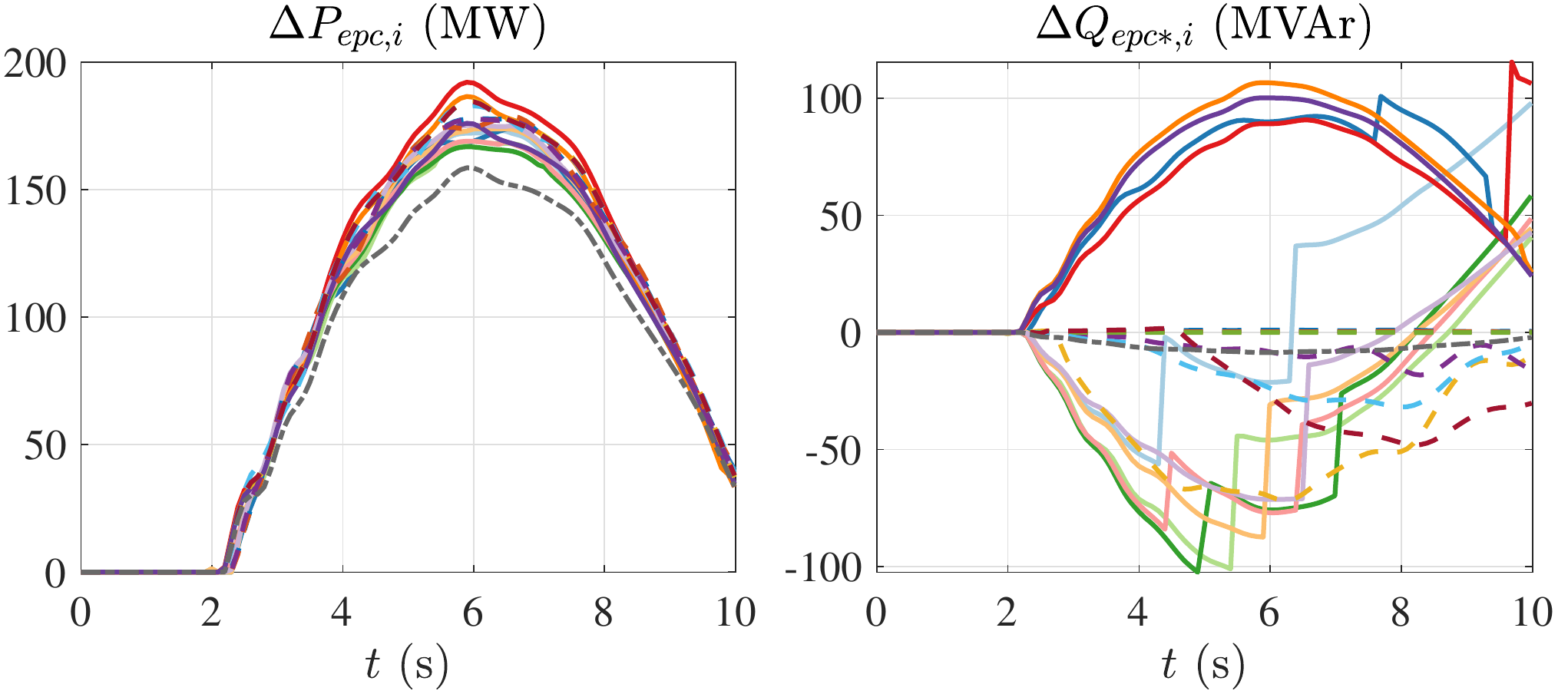}
\caption{{Difference in active and reactive power responses of HVDC links (plus LCC shunts) between EPC plus FCR-only (B) and FCR-only case (A), defined with \eqref{eq:P_delta} and \eqref{eq:Q_delta}}, respectively. See Fig.~\ref{fig:HVDClist} for legend.}
\label{fig:all_epc_tests}
\end{figure}

In further, all HVDC links which can potentially participate in EPC are assessed individually for the presented disturbance. To compare them, the focus is given only on the difference between single-EPC (B) and FCR-only (A) cases since the latter one is the same for all. Eighteen responses of HVDC links are given in Fig.~\ref{fig:all_epc_tests} (meaning, there are eighteen different simulations), which are connected to the exact coloring and notation of Fig~\ref{fig:HVDClist}. Power responses represent a difference between the applied EPC and the FCR-only case for the link which provides EPC. \footnote{There are slight changes in HVDC power response between other non-providing EPC links among different cases, but those are neglected. However, it is also discussed how the nearby VSC HVDC could interact with the EPC providing LCC HVDC links.}

\begin{figure*}[h]
\centering
\includegraphics[width=0.85\textwidth]{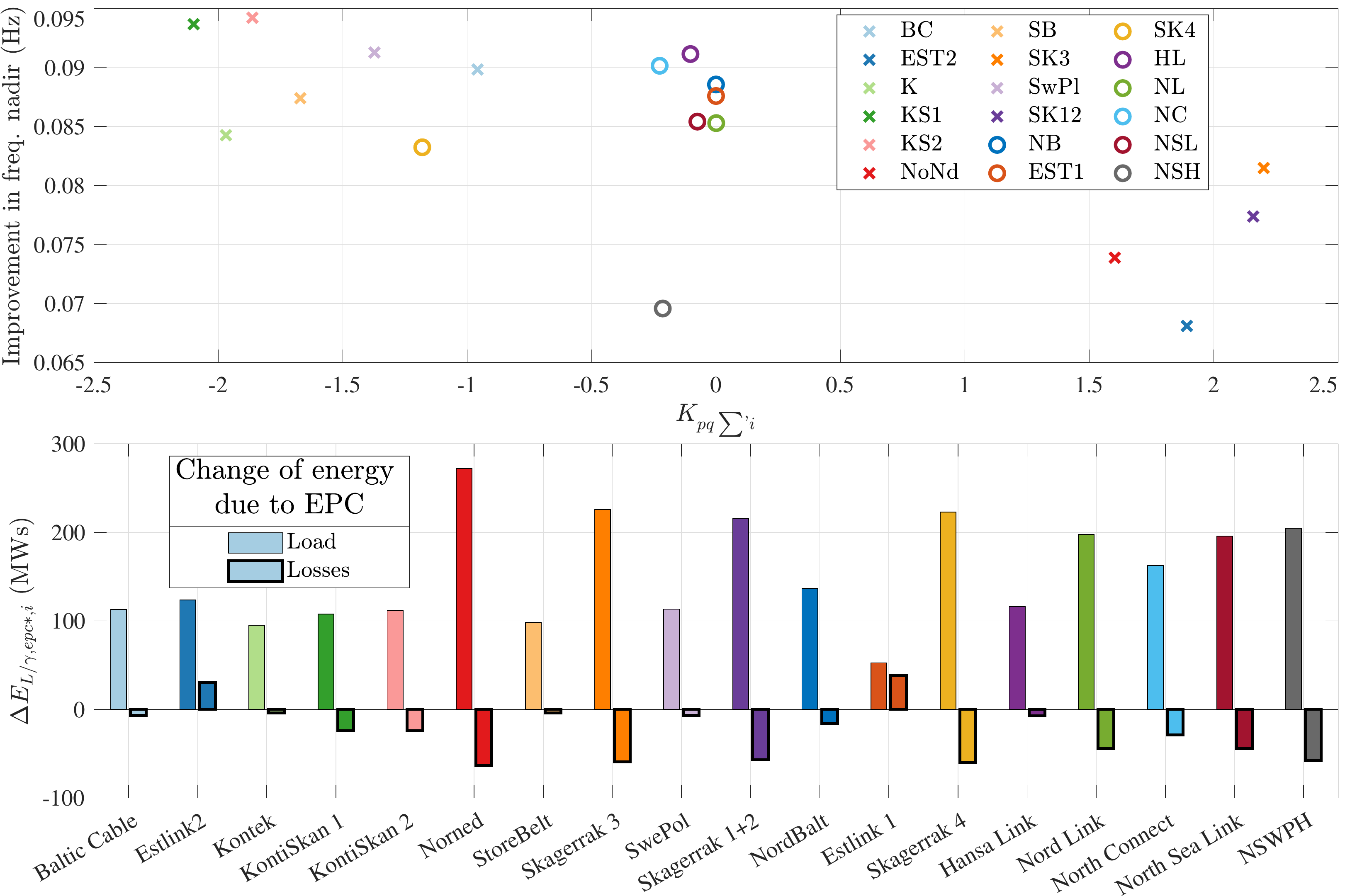}
\caption{{I) Upper figure: HVDC EPC assessment of performance to improve max IFD with respect to $\Delta K_{pq\sum,i}$ of tested link $i$, defined in \eqref{eq:Kpqsum}. II) Bottom figure: The (active) energy change of loads and losses, defined in \eqref{eq:EPL} and \eqref{eq:EPgamma}, respectively, for different cases of EPC link support.}}
\label{fig:epc_inidiv}
\end{figure*}

The applied EPC gain should not be too high and imply the violation of active power limits of any tested links, and that is satisfied by fulfilling the following condition:
\begin{equation}
    g'_i=g_i \frac{P_{r,i}}{f_n} \leq \frac{\Delta P_{\text{epc},i,max}}{f_\text{activ.,epc}-f_{max,\text{fcr}}}, i \in E,
\end{equation}

where, referring to $i^\text{th}$ link of set $E$ illustrated in Fig.~\ref{fig:HVDClist}, $\Delta P_{\text{epc},i,max}$ is the available EPC capacity in MW, and $f_{max,\text{fcr}}$ is the maximum IFD for the FCR-only case, equal to 1.226~Hz (see Fig.~\ref{fig:FCR-only}).

On the other side, the tested gain could be smaller, but in that case, there will be a marginal impact on the system response and nonlinearities of required EPC response will not be highlighted. EPC gains $g_i$ (pu) are different among each other, but $g'_i$ (MW/Hz) are the same in order to have the fair comparison. 

As explained previously, it is expected that the response of active powers are quite similar, while reactive powers behave differently. LCC HVDC links during normal operation consume power regardless of import or export scenario. 
However, consumption of reactive power has opposite behavior when EPC is applied depending on:
\begin{itemize}
    \item LCC \textit{import} plus EPC implies an \textit{increase} in reactive power consumption;
    \item LCC \textit{export} plus EPC implies a \textit{decrease} in reactive power consumption,
\end{itemize}
following the relationship defined in \eqref{eq:Q_LCC} and \eqref{eq:VdId_LCC}. It is assumed that EPC injects a positive power into the grid following an under frequency event.

%When exporting power, with EPC contribution, the absolute value of export is decreased, implying that consumed reactive power is decreased as well, following the relationship defined in \eqref{eq:Q_LCC}. On the other side, with the import scenario, both active and reactive power increase.  

That opposite behavior of reactive power of LCC HVDC links during EPC support implies different impact on terminal voltage, and nearby voltage-dependent active power load.

The reactive power response of HVDC LCC links is measured at the high-voltage transformer terminal, so plots in Figs.~\ref{fig:FCR_only_power}-~\ref{fig:all_epc_tests} consider the contributions of reactive shunts. They are activated to support the voltage profile at the LCC PCC and depend on the initial power flow scenario and rating of a converter.

The reactive power response of VSC HVDC links after a large disturbance depends on the converter capacity and parameters responsible for AC voltage control. From Fig.\ref{fig:FCR_only_power} it can be observed that some of the links reach their reactive power limits immediately after the fault, while others manage to support the AC voltage more efficiently. Naturally, if no limits are reached, VSC links reduce their reactive power injections comparing the EPC and FCR-only case, as shown in Fig.~\ref{fig:all_epc_tests}. 

\textit{Remark:} The injections of active or reactive power increase the voltage magnitude of that bus. EPC support increases the active power injection. Then, to control the voltage at the same reference, negative reactive power change is implied from the VSC voltage control loop. That is why dashed lines referring to VSC HVDC (if not saturated) are below zero in Fig.~\ref{fig:all_epc_tests}.   

Fig.~\ref{fig:epc_inidiv} shows the improvement in frequency nadir values for compared HVDC links as a function $K_{pq\sum,i}$, defined in \eqref{eq:Kpqsum}. The results of the nadir improvements can be correlated, but they are not strictly explained by $K_{pq\sum,i}$, as linearzied approach also indicated with \eqref{eq:df_ext} and \eqref{eq:df_ext2}. %%%\textcolor{red}{Nevertheless, it interesting to observe that typically, group of LCC links with positive values of $K_{pq\sum,i}$ give lower values of frequency nadir.}

To explain further the behavior of frequency response, the change of energy of loads (first bars) and losses (second bars), defined in \eqref{eq:EPL} and \eqref{eq:EPgamma}, respectively, are given in Fig.\ref{fig:epc_inidiv}. It can be seen that typically, higher improvements in max IFD are correlated to lower increase in loads and losses (their energy){---}highlighting the importance of both EPC active and side-effected reactive powers location's injection. 

%It is interesting to notice that NorNed link gives the largest overall impact on both loads and losses, even though Estlink~2 gives the lowest nadir improvement. KontiSkan 1 and 2 provide the best improvement in nadir following the high negative value of $K_{pq\sum,i}$. For the observed cases, VSC links besides Skagerrak 4 have small values of $K_{pq\sum,i}$ due to reactive current limits, and quite similar impact on the nadir. Estlink 1 and 2 increase the losses, which makes sense since they are the only ones providing the support from Finland (see Fig.~\ref{fig:sys_model}).

\textit{Observations:}
\begin{itemize}
    \item NorNed link gives the largest overall impact on the summarized loads and losses following the lower efficiency in the nadir improvement;
    \item Estlink~2 gives the lowest nadir improvement;
    \item KontiSkan 1 and 2 provide the best improvement in nadir following the high negative value of $K_{pq\sum,i}$;
    \item VSC links besides Skagerrak 4 have small values of $K_{pq\sum,i}$ due to reactive current limits, and quite similar impact on the nadir;
    \item Estlink 1 and 2 are the only ones increasing the losses, which makes sense since they are the only ones providing the support from Finland and "stress" the load flow even more (see Fig.~\ref{fig:sys_model});
    \item NSWPH HVDC link notably provides lower EPC active power output, as seen in Fig.~\ref{fig:all_epc_tests}. The main reason is its participation in power-sharing between other links in the NSWPH, as explained in \cite{NEACDCPS}. Therefore, applying the same EPC gain does not provide the same performance as it is the case for "regular" point-to-point links;
    \item Results presented in Fig.~\ref{fig:epc_inidiv} indicate that SK3 does not provide so much of less frequency improvement than SK4 even though there is a fundamental difference between $K_{pq\sum,i}$ values. The reason for that is because they are connected to the same bus. Therefore, they also interact with each other, and injection of reactive power by LCC SK3 link is compensated by the reactive power adjustments in VSC SK4 (since its reactive power is not saturated). These adjustments cannot be seen in the plot since the measurements are only given for the link which provide EPC. An alternative to overcome this might be a unique assessment of the links which are connected at the same bus or nearby.
\end{itemize}

%\subsection*{Results related to NSWPH}

%NSWPH HVDC link notably provides lower EPC active power output, as seen in Fig.~\ref{fig:all_epc_tests}. The main reason is its participation in power-sharing between other links in the NSWPH, as explained in \cite{NEACDCPS}. Therefore, applying the same EPC gain does not provide the same performance as it is the case for "regular" point-to-point links.     

%\subsection*{Results related to SK3 and SK4}

%Results presented in Fig.~\ref{fig:epc_inidiv} indicate that SK3 does not provide so much of less frequency improvement than SK4 even though there is a fundamental difference between $K_{pq\sum,i}$ values. The reason for that is because there are two HVDC links connected to the same bus. Therefore, they also interact with each other, and injection of reactive power by LCC SK3 link is compensated by the reactive power adjustments in VSC SK4. These adjustments cannot be seen in the plot since the measurements are only given for the link which provide EPC. An alternative to overcome this might be a unique assessment of the links which are connected at the same bus or nearby. 

\subsection{Results concerning various disturbances}

\textcolor{black}{The previously presented results concern the fixed disturbance, which is the most severe one according to N-1 criteria. Nevertheless, it is of interest to analyze different disturbances and if there is a change in links' EPC rankings regarding their ability to improve the maximum IFD.}

\begin{figure*}[h]
\centering
\includegraphics[width=0.85\textwidth]{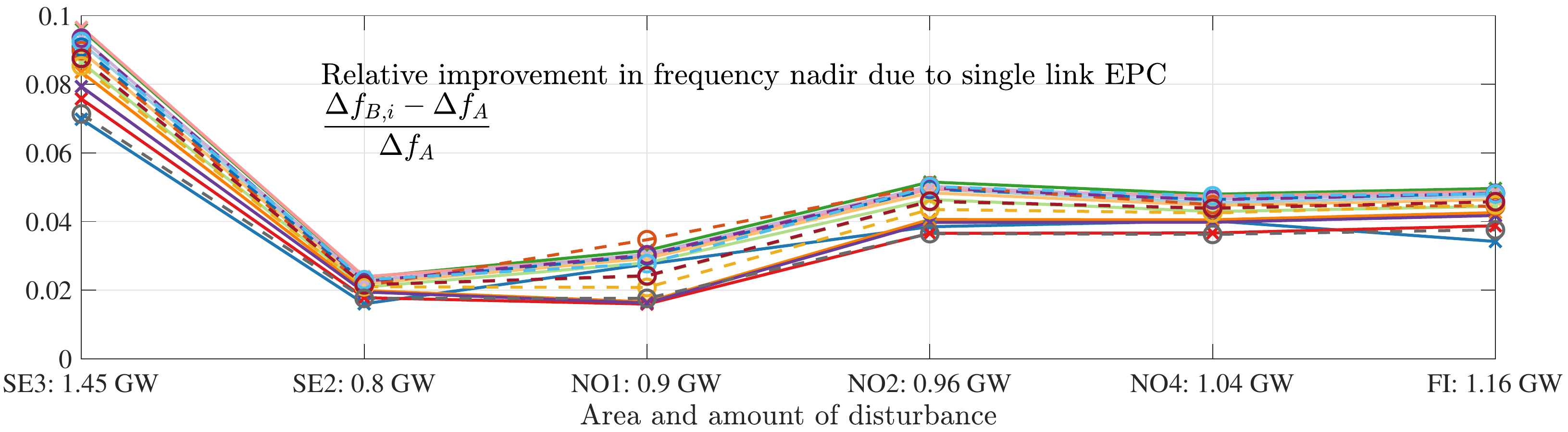}
\caption{Relative frequency improvement among HVDC links for the same EPC gain and various disturbances. See Fig.~\ref{fig:HVDClist} for legend.}
\label{fig:rel_impr}
\end{figure*}

\textcolor{black}{The results in Fig.~\ref{fig:rel_impr} illustrate the ability of HVDC EPCs to provide relative frequency nadir improvements for the different generators trips in the system. The x-axis of the figure shows the zone where disturbance is made and its active power. These generators are also marked with blue color in Fig.~\ref{fig:sys_model}.}

\textcolor{black}{The relative improvement in frequency is the largest for the dimensioning incident. Following the most significant disturbance, there is the most prominent frequency drop, so the EPC has a relatively more effective impact and opportunity to improve the frequency. Depending on the case, there could be various percentage-wise comparisons among links for the rest of the disturbances. The "NO1" and "SE2" cases are examples of the largest and smallest difference among links' EPC performance.}

\subsection{\textcolor{black}{Summarized EPC response to support the maximum IFD}}

\textcolor{black}{The previous studies were done for the specified EPC gain in MW/Hz. For the NPS, the relevant total EPC gain is the one that provides the improvement to the maximum allowed IFD of 0.9~Hz. The presented results show that value is reached with different EPC power depending on the EPC distribution.}

\textcolor{black}{Referring to Fig.~\ref{fig:epc_inidiv}, let there be two options for EPC distribution with the following MW/Hz gains:}
\begin{itemize}
    \item \textit{Option I)} (or the "bad one") NorNed: 500~MWHz, Estlink 2: 500~MW/Hz, Skagerrak 1: 125~MW/Hz, and Skagerrak 2: 125~MW/Hz;
    \item \textit{Option II)} (or the "good one") KontiSkan 1: 250~MWHz, KontiSkan 2: 250~MW/Hz, SwePol: 250~MW/Hz, North Connect: 250~MW/Hz, and Hansa Link: 250~MW/Hz.
\end{itemize}

\begin{figure}[h]
\centering
\includegraphics[width=0.48\textwidth]{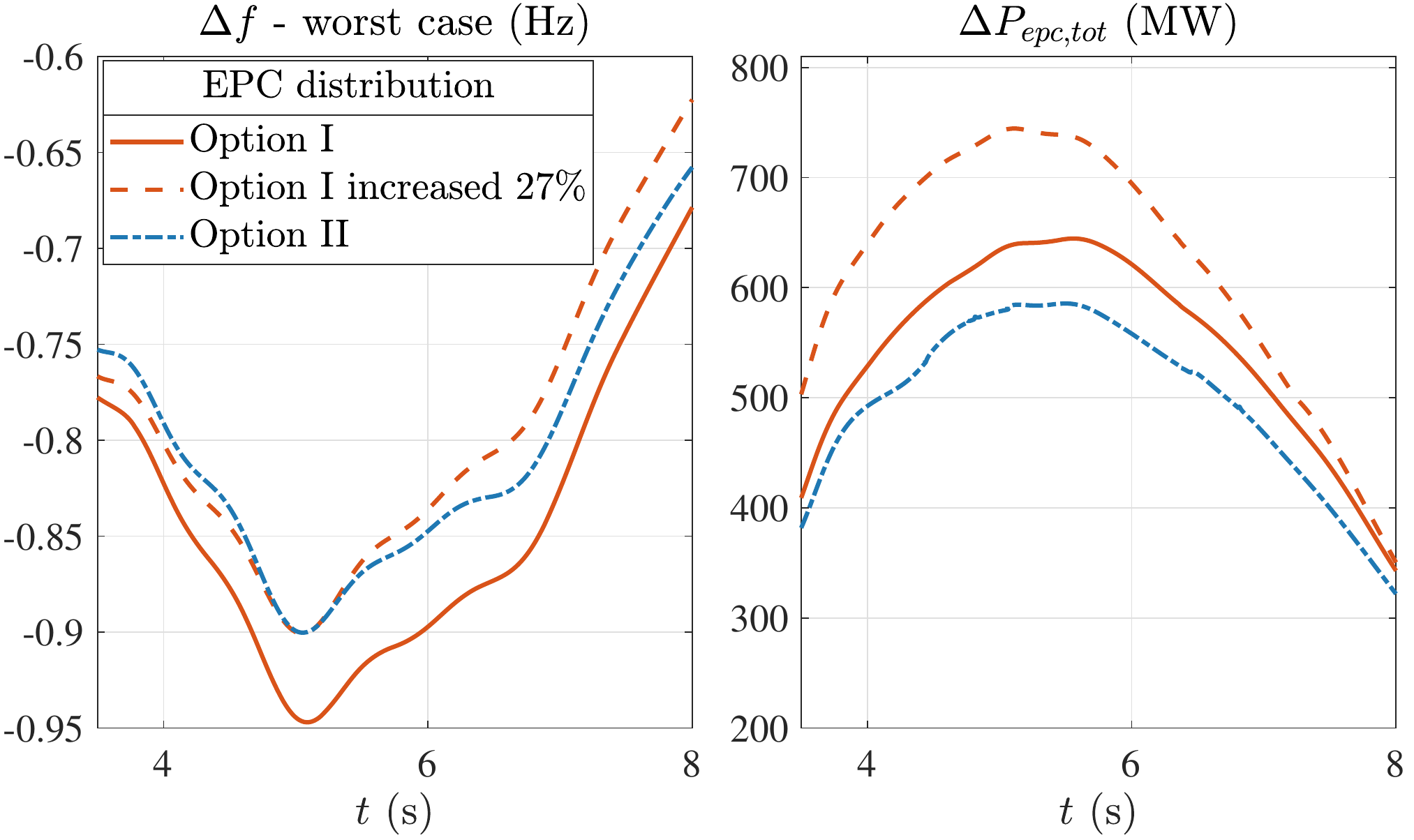}
\caption{\textcolor{black}{The comparison between EPC applied in I) SK12, EST2 and NoNd, and II) KS1, KS2, SwPl, NC, and HL with the same MW/Hz. On the left, there are the generator frequencies with the lowest nadirs for the respective cases, and on the right, total EPC contributions in MW from the listed links.}}
\label{fig:tot_epc_f}
\end{figure}

\textcolor{black}{Both options have the total EPC gain of 1250~MW/Hz by default. Their performance is shown in Fig.~\ref{fig:tot_epc_f}. Clearly, Option II performs better and can keep the frequency within the limit of 0.9~Hz of deviation, in contrast to Option I. For the EPC distribution Option I to reach 0.9~Hz, an additional increase of 27\% of EPC gain is needed (an equal increase on all EPC links), as also shown in Fig.~\ref{fig:tot_epc_f}. That implies a difference of around 160~MW between total EPC reserves on these two options for the same maximum allowed IFD for the assessed cases.}

\subsection{Main takeaways}

In the following, the study provides the more general observations made from the given results and theoretical findings.

\begin{enumerate}
    \item EPC decreases frequency deviation by affecting the active power of the utilized HVDC link. However, the question is if there is a difference in performance depending on the location, HVDC type and control design, and operating conditions.
    \item The ones that create difference are typically electrically close to large voltage-dependent loads.
    %\item This impact is dependent on their electrical distance to voltage-dependent load, disturbance, and initial power flow scenario.
    \item The EPC of HVDC LCC links may have a different ability to improve the frequency deviation after a large power disturbance depending on if they import or export power. This property implies the sign of their side-effected reactive power response due to the EPC.
    \item The contribution of LCC side-effected reactive power can be mitigated by nearby voltage controllers, such as those implemented in VSC or generators. However, that is valid only if those voltage controllers are not saturated, which can happen during a large power disturbance. 
    \item When VSC HVDC provides EPC, and its reactive power is not saturated, it tends to inject reactive power, which further decreases the frequency deviations.
    \item Each operating scenario and predefined disturbance choice define the specific list of "good" links to utilize for the EPC. They are typically "importing" LCC links, which are not electrically close to voltage controllers, and VSC, which reactive power does not get saturated during the disturbance. Also, it is preferred that power injections do not increase the losses in the grid.
    \item The difference between the "good" or "bad" distribution of EPC could imply around 25\% of the total EPC active power difference needed to fulfill the same frequency requirements.
\end{enumerate}

\section{Conclusion}

\textcolor{black}{The studies on the carefully designed NPS show that FCR fails to keep the maximum IFD within the defined margins. Therefore, additional support from the HVDC links is assessed in the form of EPC. The main question is how EPC should be distributed when the droop frequency-based control method is applied. To answer that, this study investigated the performance of each HVDC links' EPC concerning its ability to improve maximum IFD for the specified gain. The rationale to assess different HVDC links is i) injection of additional power in various locations implies a different impact on responses of total loads and losses, and ii) apart from active power change, due to control coupling in HVDC converter, there is an inherent response of reactive power (from both HVDC and shunts) further impacting the response of loads and losses.}

\textcolor{black}{Firstly, system linearization showed that the ratio between provided EPC active power and side affected reactive power correlates to the frequency response. To capture the nonlinearities of response, this study considers the integral of the given ratio from the moment of EPC activation to the moment of frequency nadir. It is shown that LCC reactive power response primarily depends on import/export scenario, while VSC on the applied AC voltage control strategy. }

\textcolor{black}{It is shown that this property (calculated integral) is well correlated with the rankings in frequency improvement among links but does not fully explain it. The strength of how much the EPC actually affects the improvement in maximum IFD is further investigated through the difference in energy of total loads and losses in the system due to applied EPC until the frequency nadir. For the given system configuration and positions of HVDC links, losses are typically decreased (besides Estlinks), while due to large EPC active power injections, the loads are increased, but to various extents.}

\textcolor{black}{The differences in EPC performances are also confirmed for various disturbances, showing that they could vary in relative impact to frequency improvement and links' comparison. }

\textcolor{black}{Finally, to present the result's importance, this work shows the needed EPC for the two opposite distribution choices, i.e., the good and bad options in terms of performance according to previously obtained results. It concludes that, for the largest disturbance and low inertia scenario, to keep the frequency within limits, the given EPCs are different by around 27\% in total EPC gain, reflecting around 160~MW in total EPC power. With further inertia reduction, it is expected for these MW only to increase and the question of EPC distribution to be more pronounced.} 

% conference papers do not normally have an appendix

% use section* for acknowledgment
\section*{Acknowledgment}

The authors would like to thank Thierry Van Cutsem, an independent consultant/adviser and a retired Adjunct Professor of the University of Li\`{e}ge and Research Director of the Fund for Scientific Research, for valuable discussions and comments on this work.

This work is supported by the multiDC project, funded by Innovation Fund Denmark, Grant Agreement No. 6154-00020B.

% trigger a \newpage just before the given reference
% number - used to balance the columns on the last page
% adjust value as needed - may need to be readjusted if
% the document is modified later
%\IEEEtriggeratref{8}
% The "triggered" command can be changed if desired:
%\IEEEtriggercmd{\enlargethispage{-5in}}

% references section

% can use a bibliography generated by BibTeX as a .bbl file
% BibTeX documentation can be easily obtained at:
% http://mirror.ctan.org/biblio/bibtex/contrib/doc/
% The IEEEtran BibTeX style support page is at:
% http://www.michaelshell.org/tex/ieeetran/bibtex/
%\bibliographystyle{IEEEtran}
% argument is your BibTeX string definitions and bibliography database(s)
%\bibliography{IEEEabrv,../bib/paper}
%
% <OR> manually copy in the resultant .bbl file
% set second argument of \begin to the number of references
% (used to reserve space for the reference number labels box)

% that's all folks
\end{document}